\documentclass[twocolumn,showpacs,preprintnumbers, amsmath,amssymb,amsfonts, final, a4paper,aps, citeautoscript,footinbib,prb,superscriptaddress]{revtex4-1}
\usepackage[pdftex]{graphicx,color}
\usepackage[latin1]{inputenc}
\usepackage{mathrsfs}
\usepackage{epstopdf}
\usepackage{times}
\usepackage{float} 
\usepackage[vcentermath]{youngtab}
\usepackage[colorlinks,bookmarks=true,citecolor=blue,linkcolor=blue,urlcolor=blue, breaklinks=true]{hyperref}
\graphicspath{{Figures/}}

\newcommand{\be}{\mathrm{e}}

\begin{document}
\title{One-dimensional two-orbital SU($N$) ultracold fermionic quantum gases 
at incommensurate filling: a low-energy approach}
\author{V. Bois}
\author{P. Fromholz}
\author{P. Lecheminant}
\affiliation{Laboratoire de Physique Th\'eorique et
 Mod\'elisation, CNRS UMR 8089,
Universit\'e de Cergy-Pontoise, Site de Saint-Martin,
F-95300 Cergy-Pontoise Cedex, France.}
\date{\today}

\begin{abstract}
We investigate the zero-temperature phase diagram of two-orbital SU($N$) fermionic models at incommensurate filling which are directly relevant to strontium and ytterbium ultracold atoms loading into a one-dimensional optical lattice.
Using a low-energy approach that takes into account explicitly the SU($N$) symmetry, we find that a spectral gap for 
the nuclear-spin degrees of freedom is formed for generic interactions. Several phases with one or two gapless
modes are then stabilized which describe the competition between different density instabilities. In stark contrast to
the $N=2$ case, no dominant pairing instabilities emerge and the leading superfluid one is rather formed from bound states of
$2N$ fermions. 
\end{abstract}
\maketitle
\section{Introduction}

Ultracold gases of alkaline-earth-like atoms have recently attracted much interest due to their
striking properties. One remarquable property is the large decoupling between electronic and nuclear
spin degrees of freedom for states with zero total electronic angular momentun 
\cite{Cazalilla-H-U-09,Gorshkov-et-al-10}. It leads to collisional properties which are independent of the nuclear
spin states and the emergence of an  extended SU($N$) continuous symmetry where $N = 2I +1$ ($I$ being
the nuclear spin) can be as large as 10 for $^{87}$Sr  fermionic atoms. 
These atoms as well as $^{171}$Yb, $^{173}$Yb ones have been cooled down to reach the quantum
degeneracy. \cite{Desalvo2010,Taie2010} In this respect,
the Mott-insulating phase, when loading these atoms in optical lattice, and the one-dimensional (1D) 
Luttinger physics of these multicomponent fermions have been investigated experimentally recently.
\cite{Taie2012,BlochFernandes2015,Pagano2014}
It paves the way of the quantum simulation of SU($N$)  many-body physics with the realization
of exotic phases such as SU($N$) chiral spin liquid states or SU($N$) symmetry-protected topological phases. 
\cite{Hermele-G-11,Hermele-G-R-09,Cazalilla-R-14,Xu-13,Nonne-M-C-L-T-13,Duivenvoorden-Q-13,Bois-C-L-M-T-15,Furusaki2014,Capponi-L-T-15,Totsuka-15,Quella2015}

A second important property of alkaline-earth-like atoms is the existence of a long-lived metastable excited state 
($^{3} P_{0}$) coupled to the ground state ($^{1} S_{0}$) via an ultranarrow doubly-forbidden transition.
This makes them ideal systems for the realization of the most precise atomic clocks \cite{Katori2011}.
The existence of these two levels might also provide new experimental realization of paradigmatic models of
Kondo and heavy-fermions physics \cite{FossFeig,Isaev2015,Kawakami2015,Zhang2015,avishai2015}
or two-orbital quantum magnetism such as the Kugel-Khomskii model \cite{KugelKhomski,Gorshkov-et-al-10}.
In the latter case, the two electronic states $|g \rangle =    \;^{1}S_{0}$ and $|e \rangle =  \; ^{3} P_{0}$ simulate the orbital degree of freedom and the spin-exchange scattering between these states has been characterized recently experimentally in fermionic $^{87}$Sr and $^{173}$Yb atoms \cite{Zhang2014,Scazza2014,Cappellini2014,Bloch2015,pagano2015}. 
Despite the fact that $g$ and $e$ states possess no electronic angular momentum, 
the tuning of interorbital interactions can be performed by exploiting the existence of an orbital Feshbach resonance
between two ytterbium atoms with different orbital and nuclear spin quantum numbers. \cite{Bloch2015,pagano2015,zhai2015}
It opens an avenue for studying the interplay between the orbital and SU($N$) nuclear spin degrees of freedom
which might lead to interesting exotic many-body physics.

In this paper, we will focus on this interplay in the special 1D case by  means of perturbative and
non-perturbative field-theoretical techniques which keep track explicitly of the non-Abelian SU($N$) symmetry of the
problem. In this respect, several two-orbital  SU($N$) fermionic lattice models with contact interactions
can be considered in the context of alkaline-earth-like atoms.\cite{Gorshkov-et-al-10,Bois-C-L-M-T-15,Kobayashi-O-O-Y-M-12,Kobayashi-O-O-Y-M-14,Capponi-L-T-15}

A first lattice model, the $g-e$ model, which is directly relevant to 
recent experiments \cite{Zhang2014,Scazza2014,Cappellini2014},
is defined from the existence of four different scattering lengths that stem from the two-body collisions with
$g$ and $e$ atomic states: \cite{Gorshkov-et-al-10}
\begin{eqnarray}
&&  \mathcal{H}_{g\text{-}e} = 
  - \sum_{m=g,e}  t_{m} \sum_{i} \sum_{\alpha=1}^{N} 
   \left(c_{m\alpha,\,i}^\dag c_{m\alpha,\,i+1}  + \text{H.c.}\right)  \nonumber \\
&& - \mu \sum_{m=g,e}  \sum_i n_{m,i}  
 + \sum_{m=g,e} \frac{U_{mm}}{2} \sum_{i} n_{m,\,i}(n_{m,\,i}-1)  \nonumber \\
&& +V \sum_i n_{g,\,i} n_{e,\,i} 
  + V_{\text{ex}}^{g\text{-}e} \sum_{i,\alpha \beta} 
  c_{g\alpha,\,i}^\dag c_{e\beta,\,i}^\dag 
  c_{g\beta ,\,i} c_{e\alpha,\,i} ,
\label{eqn:Gorshkov-Ham}
\end{eqnarray} 
where $c_{m\alpha,\,i}^\dag$ denotes the fermionic creation operator on the site $i$ with nuclear spin 
index $\alpha$ ($\alpha=1,\ldots, N$ with $N=2I+1$)  and orbital index $m=g,e$  which labels the two atomic states 
${}^{1}S_{0}$ and ${}^{3}P_{0}$, respectively.  In Eq. (\ref{eqn:Gorshkov-Ham}), the local fermion numbers of the species $m=g,e$ at the site $i$ are defined by: $n_{m,i} = \sum_{\alpha=1}^{N}c^{\dagger}_{m\alpha,i}c_{m\alpha,i}$.
The $g-e$ model (\ref{eqn:Gorshkov-Ham}) is invariant under continuous U(1)$_{\text{c}}$  and SU($N$)$_{\text{s}}$ symmetries:
\begin{equation}
c_{m\alpha,\,i} \mapsto \be^{i \theta} c_{m\alpha,\,i} \, , \; 
c_{m\alpha,\,i} \mapsto \sum_{\beta} U_{\alpha \beta}  c_{m\beta,\,i} ,
\label{eqn:U(N)}
\end{equation}
$U$ being an SU($N$) matrix. The two transformations (\ref{eqn:U(N)}) 
respectively refer to the conservation of the total number of atoms, that will be called U(1)$_{c}$ charge symmetry 
in the following for simplicity, and the 
SU($N$)$_{s}$ symmetry in the nuclear-spin sector.
On top of these obvious symmetries, the Hamiltonian is also invariant under an U(1)$_{o}$ orbital symmetry:
\begin{equation}
c_{g\alpha,\,i} \mapsto \be^{i \theta_{\text{o}}} c_{g\alpha,\,i} \, , \; 
c_{e\alpha,\,i} \mapsto \be^{-i \theta_{\text{o}}} c_{e\alpha,\,i}  \; ,
\label{eqn:orbital-U1}
\end{equation}
which means that the total fermion numbers for $g$ and $e$ states are 
conserved separately in Eq. (\ref{eqn:Gorshkov-Ham}).

A second  two-orbital  SU($N$) fermionic lattice model can be defined by considering only 
the atoms in the $g$ state as in experiments \cite{Pagano2014}
and the orbital degrees of freedom are the two degenerate first-excited $p_x$ and $p_y$ states of an 2D harmonic trap.
More specifically,  alkaline-earth atoms are loaded into an 1D optical lattice (running in the $z$-direction) with moderate strength of harmonic confining potential
$m \omega^2 (x^2 +y^2)/2$ in the direction perpendicular to the chain.  
It is assumed that the $s$-level of the oscillator is fully occupied while the $p$-levels of the oscillator are partially filled.
The resulting lattice model reads as follows in the tight-binding approximation:
\cite{Kobayashi-O-O-Y-M-12,Kobayashi-O-O-Y-M-14,Bois-C-L-M-T-15} 
\begin{eqnarray}
\!\!\!&&  \mathcal{H}_{p\text{-band}}  = -t \sum_{i,m \alpha} \left(c_{m\alpha,\,i}^\dag c_{m\alpha,\,i+1}+  H.c.\right)
  -\mu\sum_i n_i \nonumber \\
\!\!\! &+& \frac{U_1 + U_2}{4} \sum_i  n_i^2
 \! +\! \sum_i\!\big[2U_2 (T_i^x)^2 + (U_1\!-\!U_2)(T_i^z)^2\big] ,
  \label{pbandmodel}
\end{eqnarray}
where 
$c^{\dagger}_{m\alpha,\,i}$  is a creation fermionic operator 
with orbital index $m=p_x, p_y$ and nuclear spin components $\alpha=1,\cdots,N$ on the $i$th site of the optical lattice. 
In Eq. (\ref{pbandmodel}),  $n_i =  \sum_{m\alpha} 
c_{m \alpha,\,i}^\dag c_{m \alpha,\,i}$ describes the density operator on the $i$th site
and a pseudo-spin operator for the orbital degrees of freedom has been defined
\begin{eqnarray}
  T_i^a&=&\frac{1}{2}c_{m \alpha,\,i}^\dag \sigma^a_{m n} c_{n \alpha,\,i} ,
  \label{defpeoperator}
\end{eqnarray}
$\sigma^a (a =x,y,z)$ being the Pauli matrices and a summation over repeated indices (except thoses which label
the lattice sites) is implied in the following.
In stark contrast to the $g-e$ model (\ref{eqn:Gorshkov-Ham}), the $p$-band model (\ref{pbandmodel}) has a hopping term
which does not depend on the orbital state. The use of harmonic potential in the $xy$ direction implies a constraint of the two coupling constants $U_{1,2}$ in Eq. (\ref{pbandmodel}): $U_1 = 3 U_2$ \cite{Bois-C-L-M-T-15}.
As we will see later, the harmonic line plays a special role for $N>2$ as the result of the competition
between several different instabilities. It is then interesting to consider the generalized $p$-band model (\ref{pbandmodel}),
which can be realized by introducing a quartic confinement potential, to fully reveal
the physics along the harmonic line. As it can be easily seen, the $p$-band model (\ref{pbandmodel}) enjoys an $\text{U}(1)_c\times\text{SU}(N)_s$  continuous symmetry which is defined by Eq. (\ref{eqn:U(N)}). Along the harmonic line  $U_1 = 3 U_2$, it displays an additional U(1)$_{o}$ symmetry which is a rotation along the y-axis in the orbital subspace. 
Model  (\ref{pbandmodel}) exhibits also this extended U(1)$_{o}$ symmetry when
$U_2=0$ or $U_1=U_2$ where it becomes equivalent to two decoupled single-orbital 
SU($N$) Hubbard chain model. In remaining cases, the U(1)$_{o}$ continuous symmetry in the orbital sector is explicitly broken in contrast to the $g-e$ model.

In this paper, we investigate the low-energy properties of the two-orbital SU($N$) models 
(\ref{eqn:Gorshkov-Ham}, \ref{pbandmodel}) at incommensurate filling by means of one-loop renormalization  group (RG) approach and non-Abelian bosonization techniques \cite{Gogolin-N-T-book,Witten84,Knizhnik-Z-84,affleck86}. 
The latter approach is crucial to fully take into account the presence of the SU($N$) symmetry
in this problem of alkaline-earth cold atoms. The half-filled case of these models has already been analysed by
complementary techniques in Refs. \onlinecite{Nonne-M-C-L-T-13,Bois-C-L-M-T-15,Totsuka-15}. Several interesting phases, 
including symmetry-protected topologically phases, have been found due to the interplay between orbital and
nuclear spin degrees of freedom. The specific $N=2$ case at incommensurate filling has been recently studied
in  Ref. \onlinecite{Bois2015} where the competition between different dominant superconducting pairing instabilities
has been revealed. As we will show here, the physics of two-orbital SU($N$) models turns out to be very different when $N>2$. We find that the zero-temperature phase diagram of these models is characterized by competiting
density instabilities. In stark contrast to the conclusion of Ref. \onlinecite{Szirmai2013},  no dominant superconducting pairing instabilities can appear in an SU($N$) spin-gap phase when $N>2$ 
since they are not singlet under the SU($N$) symmetry when $N>2$. In this respect,
we find that the leading superfluid instability is rather formed from bound states of $2N$ fermions giving rise to a molecular
Luttinger liquid behavior at sufficiently small density. \cite{Capponi2008,Roux-C-L-A-09}

The rest of the paper is organized as follows. In Sec. II, we perform the continuum limit of the two models 
(\ref{eqn:Gorshkov-Ham}, \ref{pbandmodel}) in terms of $2N$ left-right moving Dirac fermions. The effective low-energy Hamiltonian is then described in a basis where the SU($N$)  nuclear-spin symmetry is made explicit. The one-loop 
RG analysis of the continuum model is presented in Sec. III. We then map out in Sec. IV
the zero-temperature phase diagram of models (\ref{eqn:Gorshkov-Ham}, \ref{pbandmodel}). Finally, Sec. V contains our concluding remarks.  

\section{Continuum limit}\label{sec:lowenergy}
In this section, the continuum limit of the two-orbital SU($N$) Hubbard models 
(\ref{eqn:Gorshkov-Ham}, \ref{pbandmodel}) at incommensurate filling is determined.

\subsection{$p$-band model case}

Let us first consider the weak-coupling approach to the $p$-band model (\ref{pbandmodel}) 
at incommensurate filling. The latter model has two degenerate Fermi points $\pm k_F$. 
The starting point of the continuum-limit procedure is the linearization of the non-interacting energy spectrum in the vicinity
of the Fermi points and the introduction of $2N$ left-right moving Dirac fermions \cite{Giamarchi-book,Gogolin-N-T-book}:
\begin{equation}
c_{m \alpha,\,i} \rightarrow \sqrt{a_0} (L_{m \alpha}
e^{-i k_F x} + R_{m \alpha} e^{i k_F x} ),
 \label{contlimitdirac}
\end{equation}
with  $m=p_x, p_y$, $\alpha= 1, \ldots, N$, and $x =i a_0$, $a_0$ being the lattice spacing.
The non-interacting Hamiltonian density is equivalent to that of $2N$ left-right moving Dirac fermions:
\begin{equation}
  {\cal H}_0=-i v_\text{F} \left(:R_{m \alpha} ^\dag \partial_x R_{m \alpha} ^{\phantom \dag}: - 
  :L_{m \alpha}^\dag \partial_x L_{m \alpha}^{\phantom \dag}:\right) ,
\label{HamcontDirac}  
\end{equation}  
where $v_F = 2t a_0 \sin (k_F a_0)$ is the Fermi velocity and $:A:$ denotes 
the standard normal ordering of an operator $A$. The continuum limit of the $p$-band model is then achieved by replacing (\ref{contlimitdirac}) into the interacting part of the lattice model Hamiltonian (\ref{pbandmodel}) and keeping only non-oscillating contributions.

At incommensurate filling, there is no umklapp process which couples 
charge and other orbital or SU($N$) degrees of freedom. Model (\ref{pbandmodel})  enjoys then a ''spin-charge'' separation in the low-energy limit which is the hallmark of 1D conductors  \cite{Giamarchi-book,Gogolin-N-T-book}:
\begin{equation}
{\cal H}_{p\text{-band}}  = {\cal H}_c + {\cal H}_{\text{so}},
\label{spinchargesep}
\end{equation}
with $[{\cal H}_c, {\cal H}_{\text{so}}] = 0$. The physical properties of the charge degrees
of freedom are governed by ${\cal H}_c$ and  ${\cal H}_{\text{so}}$ describes the interplay between
SU($N$) nuclear spins and orbital degrees of freedom.

The charge Hamiltonian takes the form of a Tomonaga-Luttinger model with Hamiltonian density:
\begin{equation}
{\cal H}_c = \frac{v_c}{2} \left[\frac{1}{K_c} 
\left(\partial_x \Phi_c \right)^2 + K_c 
\left(\partial_x \Theta_c \right)^2 \right],
\label{luttbis}
\end{equation}
which accounts for metallic properties in the
Luttinger liquid universality class \cite{Giamarchi-book,Gogolin-N-T-book}.
In this low-energy approach, the charge excitations are described by the bosonic field $\Phi_c$ and its dual field $\Theta_c$.
The explicit form of the Luttinger parameters $v_c$ and $K_c$  in the weak-coupling regime can be extracted from the continuum limit and we find:
\begin{eqnarray}
K_c &=&  \frac{1}{\sqrt{1 + g_c/\pi v_F}}
\nonumber \\
v_c &=& v_F\sqrt{1 + g_c/\pi v_F} ,
\label{Luttingerparameters}
\end{eqnarray}
with $g_c = a_0 (N-1) (U_1 + U_2)$.

We now consider the continuum description of the Hamiltonian ${\cal H}_{\text{so}}$ in Eq. (\ref{spinchargesep}).
To this end, we need to introduce several chiral fermionic bilinear terms 
which will be useful to perform a one-loop RG analysis of ${\cal H}_{\text{so}}$. 
These quantities can be identified by exploiting the continuous symmetry of the lattice model (\ref{pbandmodel}).
The non-interacting model \eqref{HamcontDirac} enjoys an U(2$N$)$|_\text{L}$ $\otimes$ U(2$N$)$|_\text{R}$  symmetry which results from its invariance under independent unitary transformations on the $2N$ left and right Dirac fermions. Its massless properties are then governed by a conformal field theory (CFT)  U(2$N$)$_1$= U(1)$_\text{c} \times$ 
SU($2N$)$_1$ based on this U(2$N$) symmetry. \cite{affleck86} Since the $p$-band model displays an extended global SU($N$) symmetry, we need to decompose the SU(2$N$)$_1$ CFT, with $2N-1$ bosonic gapless modes, 
into a CFT which is directly related to the SU($N$) symmetry. The resulting decomposition is similar to the one
which occurs in the multichannel Kondo problem with the use of the conformal embedding  \cite{Affleck-L-91}:
U(2$N$)$_1$  $\supset$ U(1)$_c$ $\times$ SU($N$)$_2$ $\times$ SU($2$)$_N$.
In this respect, we introduce the currents which generate 
the SU($N$)$_2$ $\times$ SU($2$)$_N$ CFT of the problem:
\begin{equation}
\begin{split}
& J_\text{L}^a = L_{n\alpha}^\dagger T^a_{\alpha \beta} L_{n\beta} \quad \text{SU($N$)}_2 
\text{ (nuclear) spin currents}\\
&  j_{\text{L}}^i = \frac{1}{2} L_{m\alpha}^\dagger \sigma^i_{m n} L_{n\alpha} \quad
\text{SU(2)}_N \text{ orbital currents}  \\
&  J_{\text{L}}^{a,i} =  
 \frac{1}{\sqrt{2}}  L_{m\alpha}^\dagger T^a_{\alpha \beta} \sigma^i_{m n} L_{n\beta} \quad
\text{mixed currents},
\end{split}
\label{defalkacurrents}
\end{equation}
where
$\sigma^i$ ($i=x,y,z$)  and $T^a$ ($a= 1, \ldots, N^2 -1$) are respectively the Pauli matrices 
and SU($N$) generators in the fundamental representation of SU($N$) normalized such that: $\text{Tr}(T^a T^b)=\delta^{a\,b}/2$. The combination ${\cal J}_L^A = (J_\text{L}^a,  j_{\text{L}}^i , J_{\text{L}}^{a,i})$ with $A = 1, \ldots, 4 N^2 -1$
defines SU(2$N$)$_1$ left currents of the non-interacting theory (\ref{HamcontDirac}).

With all these definitions at hand, we are able to derive the continuum limit of the $p$-band 
model (\ref{pbandmodel}). We will neglect all chiral contributions which account for
velocity anisotropies and focus on the interacting Hamiltonian. 
One can then derive the continuum limit of  ${\cal H}_{\text{so}}$  in terms of the currents (\ref{defalkacurrents})
and similar expressions for the right-moving ones. After standard calculations, we get the interacting
part of ${\cal H}_{\text{so}}$:
\begin{eqnarray}
  &{\cal H}_{\text{so}}^{\rm int} &=
    g_1 J_L^aJ_R^a + g_2  J_L^{a,x} J_R^{a,x}
    + g_3 J_L^{a,z} J_R^{a,z}  + g_4  J_L^{a,y} J_R^{a,y} \nonumber\\
  &&
  + g_5  j_L^x j_R^x + g_6 j_L^zj_R^z  +  g_7 j_L^y j_R^y ,
    \label{lowenergyham}
\end{eqnarray}
with the following identification for the coupling constants:
\begin{equation}
\begin{array}{lll}
	g_1&=& - a_0 (U_2+U_1)   \\
	g_2&=&   - 4 a_0 U_2   \\
	g_3 &=& 2a_0 (U_2 -U_1) \\
	g_4 &=& g_7 = 0 \\
	g_5&=&  a_0  \frac{4(N-1)}{N} U_2  \\
	g_6 &=& a_0  \frac{2(N-1)}{N} (U_1 - U_2)  .
	\label{couplings}
\end{array}
\end{equation}
In the interacting Hamiltonian (\ref{lowenergyham}), we have included additional perturbations with
coupling constants $g_{4,7}$ which will be generated at one-loop order within the RG approach as we will see in Sec. III.

\subsection{$g-e$ model case}

We now turn to the continuum limit of the $g-e$ model (\ref{eqn:Gorshkov-Ham}) 
with $t_g = t_e$ and $U_{gg} = U_{ee} $ so that the only two Fermi points $\pm k_F$ we get do not depend on the orbital state as in the $p$-band model, see Eq. (\ref{contlimitdirac}). 
The continuum Hamiltonian of the $g-e$ model (\ref{eqn:Gorshkov-Ham}) at incommensurate filling 
also takes the general form (\ref{spinchargesep}) where the charge degrees of freedom are governed by the 
Tomonaga-Luttinger Hamiltonian density (\ref{luttbis}). The Luttinger parameters 
are still given by Eq. (\ref{Luttingerparameters})
with $g_c =  a_0 \left( \left( N - 1 \right) U - V_{\text{ex}}^{g\text{-}e}  + NV\right) $.

The low-energy Hamiltonian ${\cal H}_{\text{so}}$, which accounts for the interaction between orbital and nuclear
spin degrees of freedom, displays the same structure as in Eq. (\ref{lowenergyham}). The main difference stems from
the fact that we have now the constraints $g_2 = g_4$ and $g_5 = g_7$  in Eq. (\ref{lowenergyham}) 
since the $g-e$ model (\ref{eqn:Gorshkov-Ham}) is invariant under the U(1)$_{o}$ orbital symmetry.
After straightforward calculations, we get the following identification for the coupling constants:
\begin{equation}
\begin{array}{lll}
	g_1&=& -a_0 \left(U +  V_{\text{ex}}^{g\text{-}e} \right)    \\
	g_2&=&  g_4 =  -2a_0V    \\
	g_3 &=&  - 2 a_0  \left(U -    V_{\text{ex}}^{g\text{-}e}  \right)  \\
	g_5&=&  g_7= 2a_0\left( V_{\text{ex}}^{g\text{-}e} - V/N \right)  \\
	g_6 &=& \frac{2 a_0}{N} \left( \left( N - 1 \right) U + V_{\text{ex}}^{g\text{-}e}  -  NV\right)   .
	\label{couplingsgeisotrop}
\end{array}
\end{equation}

The anisotropic case  $t_g \ne t_e$ and $U_{gg} \ne U_{ee}$ of the $g-e$ model (\ref{eqn:Gorshkov-Ham})
is directly related to ytterbium and strontium atoms since the scattering lengths corresponding to the collisions between $g-g$ and $e-e$ states are different experimentally. \cite{Zhang2014,Scazza2014,Cappellini2014}
In this anisotropic case, the non-interacting energy spectrum can have now
four different Fermi points $\pm k_{g,eF}$ which depend on the orbital state. 
The linearization of the spectrum around these Fermi points is then described by:
\begin{equation}
c_{m \alpha,\,i} \rightarrow \sqrt{a_0} (L_{m \alpha}
e^{-i k_{mF} x} + R_{m \alpha} e^{i k_{mF} x} ),
 \label{contlimitdiracge}
\end{equation}
with  $m=g, e$ and $\alpha= 1, \ldots, N$.  
One important modification of the continuum-limit procedure is the impossibility to
use the basis (\ref{defalkacurrents}) of the low-energy approach which assumes an $g \leftrightarrow e$ invariance between the orbital states to define the SU($N$)$_2$ current $J^{a}_{R,L}$. 
In this respect, we introduce U(1) and SU($N$)$_1$ left-moving currents for each orbital state $m=g,e$:
\begin{eqnarray}
J_{mL} &=&   :L_{m \alpha}^\dagger L_{m\alpha}: \; = \sqrt{\frac{N}{\pi}} \partial_x \Phi_{mL} \nonumber \\
J^{a}_{mL} &=&  L_{m \alpha}^\dagger T^{a}_{\alpha \beta} L_{m\beta}  ,
\label{currentaniso}
\end{eqnarray}
with $a= 1, \ldots, N^2 -1$ and similar definitions for the right-movers. 
The bosonic field $\Phi_{m} = \Phi_{mL} + \Phi_{mR}$ describes the U(1)$_m$ 
density fluctuations in the orbital $m=g,e$ subspace.

At incommensurate filling, the continuum Hamiltonian separates into the U(1)$_g$ $\times$ U(1)$_e$ part which
stands for $g$ and $e$ orbital density fluctuations,  and a non-Abelian part corresponding to the symmetry breaking 
SU($N$)$_g$ $\times$ SU($N$)$_e$ $\rightarrow$ SU($N$)$_s$ when the interactions are switched on. 
The Abelian part can be expressed in terms of the bosonic fields for each orbital:
\begin{eqnarray}
 {\cal H}_{U(1) \times U(1)} &=& \sum_{m=g,e} \frac{v_{m}}{2} 
 \left[ \frac{1}{K_m} 
 \left(\partial_x \Phi_m \right)^2 + K_m \left(\partial_x \Theta_m \right)^2 \right]
 \nonumber \\
 &+&  \frac{a_0(N V - V_{\text{ex}}^{g\text{-}e} )}{\pi} \partial_x \Phi_g \partial_x \Phi_e,
\label{HamU1contge}
\end{eqnarray}
where $\Theta_m = \Phi_{mL} - \Phi_{mR}$ is the dual field to $\Phi_{m}$ and 
the Luttinger parameters are:
\begin{eqnarray}
K_m &=&  \left(1 + \frac{a_0 (N-1)U_{mm}}{\pi v_{mF}}\right)^{-1/2}
\nonumber \\
v_m &=& v_{mF}\left(1 + \frac{a_0 (N-1)U_{mm}}{\pi v_{mF}}\right)^{1/2} ,
\label{Luttingerparametersmorb}
\end{eqnarray}
$v_{mF} = 2 t a_0 \sin(k_{Fm} a_0) $ being the Fermi velocity for the band $m=g,e$.

The Hamiltonian (\ref{HamU1contge}) can then be diagonalized by introducing
the charge and orbital fields: \cite{Szirmai2013}
\begin{eqnarray}
 \Phi_c &=& \frac{1}{\sqrt{v_g +v_e}} \left( \sqrt{\frac{v_g}{K_g}} \; \Phi_g +  \sqrt{\frac{v_e}{K_e}} \; \Phi_e \right)
 \nonumber \\
\Phi_o &=& \frac{1}{\sqrt{v_g +v_e}} \left( \sqrt{\frac{v_g}{K_g}} \; \Phi_g -  \sqrt{\frac{v_e}{K_e}} \; \Phi_e \right),
\label{transU1}
\end{eqnarray}
so that it takes the form of two decoupled Tomonaga-Luttinger liquid Hamiltonian
densities:
\begin{eqnarray}
{\cal H}_{U(1) \times U(1)} &=& \frac{v_c}{2} \left[\frac{1}{K_c} 
\left(\partial_x \Phi_c \right)^2 + K_c 
\left(\partial_x \Theta_c \right)^2 \right] \nonumber \\
 &+& \frac{v_o}{2} \left[\frac{1}{K_o} 
\left(\partial_x \Phi_o \right)^2 + K_o 
\left(\partial_x \Theta_o \right)^2 \right] ,
\label{HamU1}
\end{eqnarray}
where $K_c$ and $K_o$ (respectively $v_c$ and $v_o$) 
are Luttinger parameters (respectively velocities) 
for respectively the charge and orbital degrees of freedom:
\begin{eqnarray}
K_{c,o} &=&  \left(1 \pm \frac{a_0 (NV - V_{\text{ex}}^{g\text{-}e})\sqrt{K_g K_e}}{\pi  \sqrt{v_g v_e}}\right)^{-1/2}
\nonumber \\
v_{c,o} &=& {\bar v} \left(1 \pm \frac{a_0 (NV - V_{\text{ex}}^{g\text{-}e})\sqrt{K_g K_e}}{\pi  \sqrt{v_g v_e}}\right)^{1/2} ,
\label{Luttingerparametersco}
\end{eqnarray}
$ {\bar v} = (v_g + v_e)/2$ being the average velocity.

The interacting part of the non-Abelian sector takes 
the form of an SU($N$)$_1$  current-current interaction after neglecting 
chiral contributions as before:
\begin{eqnarray}
\!\!\!\!\!{\cal H}^{\rm int}_{\rm SU(N)} \!=\!\!\! \sum_{m=g,e} \!\!\lambda_m J^{a}_{mL} J^{a}_{mR} 
+  \lambda \!  \left( J^{a}_{g L} J^{a}_{ e R}  \!+\! J^{a}_{e L} J^{a}_{g R}  \right) ,
\label{HamSUN}
\end{eqnarray}
with $ \lambda_{g} = - 2  U_{gg} a_0$, $ \lambda_{e} = - 2  U_{ee} a_0$, and $ \lambda = - 2V_{\text{ex}}^{g\text{-}e} a_0$.

\section{One-loop renormalization group analysis}\label{sec:contlimit}
The one-loop RG calculation enables one to deduce the infrared (IR) properties of  
model (\ref{lowenergyham}) and thus to map out
the zero-temperature phase diagram of the  $g-e$ model (\ref{eqn:Gorshkov-Ham}) and $p$-band  model (\ref{pbandmodel}).

\subsection{RG approach for the $p$-band model}
We start with the $p$-band model case.
To this end, it is convenient for the RG analysis to introduce the following rescaled coupling constants:
\begin{eqnarray}
g_1 ~ &=& \phantom{+}\frac{\pi}{\mathrm{N}}f_1\nonumber \\
g_{2,3,4} &=& \frac{2\pi}{\mathrm{N}}f_{2,3,4}\nonumber \\
g_{5,6,7} &=& \frac{2\pi}{\mathrm{N}^2}f_{5,6,7}.
\end{eqnarray}
After cumbersome calculations, we find the following one-loop RG equations:
\begin{eqnarray}
\dot{f}_{1} &=& \frac{1}{4}\left(f_1^2+f_2^2+f_3^2+f_4^2\right) \nonumber \\
\dot{f}_{2} &=& \frac{1}{2}f_1f_2 + \frac{\mathrm{N}^2-4}{2\mathrm{N}^2}f_3f_4 + \frac{1}{\mathrm{N}^2}\left(f_4f_6 + f_3f_7\right) \nonumber \\
\dot{f}_{3} &=& \frac{1}{2}f_1f_3 + \frac{\mathrm{N}^2-4}{2\mathrm{N}^2}f_2f_4 + \frac{1}{\mathrm{N}^2}\left(f_2f_7 + f_4f_5\right) \nonumber \\
\dot{f}_4 &=& \frac{1}{2}f_1f_4 + \frac{\mathrm{N}^2-4}{2\mathrm{N}^2}f_2f_3 + \frac{1}{\mathrm{N}^2}\left(f_2f_6 + f_3f_5\right) \nonumber \\
\dot{f}_{5} &=& \frac{\mathrm{N}^2-1}{\mathrm{N}^2}f_3f_4 + \frac{1}{\mathrm{N}^2}f_6f_7 \nonumber \\
\dot{f}_{6} &=& \frac{\mathrm{N}^2-1}{\mathrm{N}^2}f_2f_4 + \frac{1}{\mathrm{N}^2}f_5f_7 \nonumber \\
\dot{f}_7 &=& \frac{\mathrm{N}^2-1}{\mathrm{N}^2}f_2f_3 + \frac{1}{\mathrm{N}^2}f_5f_6 .
\label{RGbis}
\end{eqnarray}

\begin{figure}[!ht]
\centering
\includegraphics[width=0.5\columnwidth,clip]{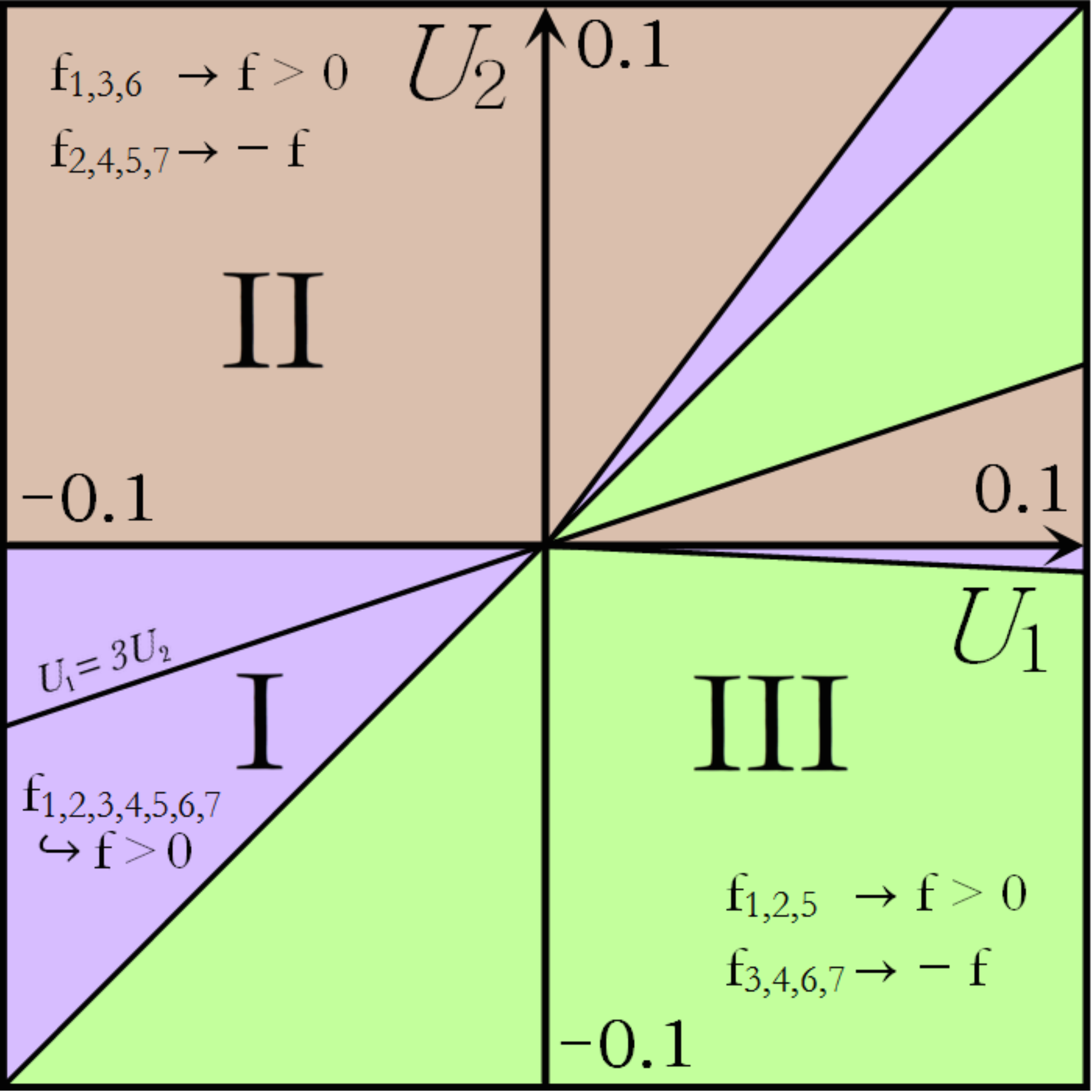}
\caption{Numerical phase diagram  of the $p$-band model (\ref{pbandmodel}) obtained from the one-loop RG analysis for $N>2$. }
\label{fig:PhasediagNRG}
\end{figure}

The analysis of the RG Eqs. (\ref{RGbis}) for $N=2$ has been done in Ref. \onlinecite{Bois2015}. 
The latter case is very special since the one-loop RG Eqs. (\ref{RGbis}) contains several terms which vanish.
When $N>2$ case, Eqs. (\ref{RGbis}) have less symmetries and the analysis is more involved.
We performed the numerical analysis of the RG Eqs. (\ref{RGbis}) by a Runge-Kutta procedure. 
The RG flow in the $N=3$ case turns out to be very slow. In this respect, the vicinity of the transition lines of Fig.  \ref{fig:PhasediagNRG} in the repulsive regime are difficult to be resolved numerically for $N=3$. When $N>3$, the analysis
is much simpler and it leads to the results presented in Fig. \ref{fig:PhasediagNRG}.

As depicted in Fig. \ref{fig:PhasediagNRG}, our numerical results reveal that
the RG flow goes in the strong-coupling regime in the far IR along three asymptotic lines:
\begin{eqnarray}
\!\!\!\!\!\!\!\!&\mbox{I}& \; : f_2 = f_3 = f_4= f_5 =  f_6 = f_7= f_1 = f^{*} > 0\nonumber \\
\!\!\!\!\!\!\!\!&\mbox{II}& \; : f_2 = - f_3 = f_4 =  f_5 =   - f_6  =  f_7 = - f_1 = -  f^{*}\nonumber \\
\!\!\!\!\!\!\!\!&\mbox{III}& \; : f_2 = - f_3 = - f_4 = f_5  =  - f_6  =  - f_7 = f_1 = f^{*} .
\label{asymptotRG}
\end{eqnarray}
These lines give rise to the three different regions of Fig. \ref{fig:PhasediagNRG} that we analyse in
the next section.

\subsection{RG approach for the $g-e$ model}\label{sec:RGge}
 
\begin{figure}[!ht]
\centering
\includegraphics[width=\columnwidth,clip]{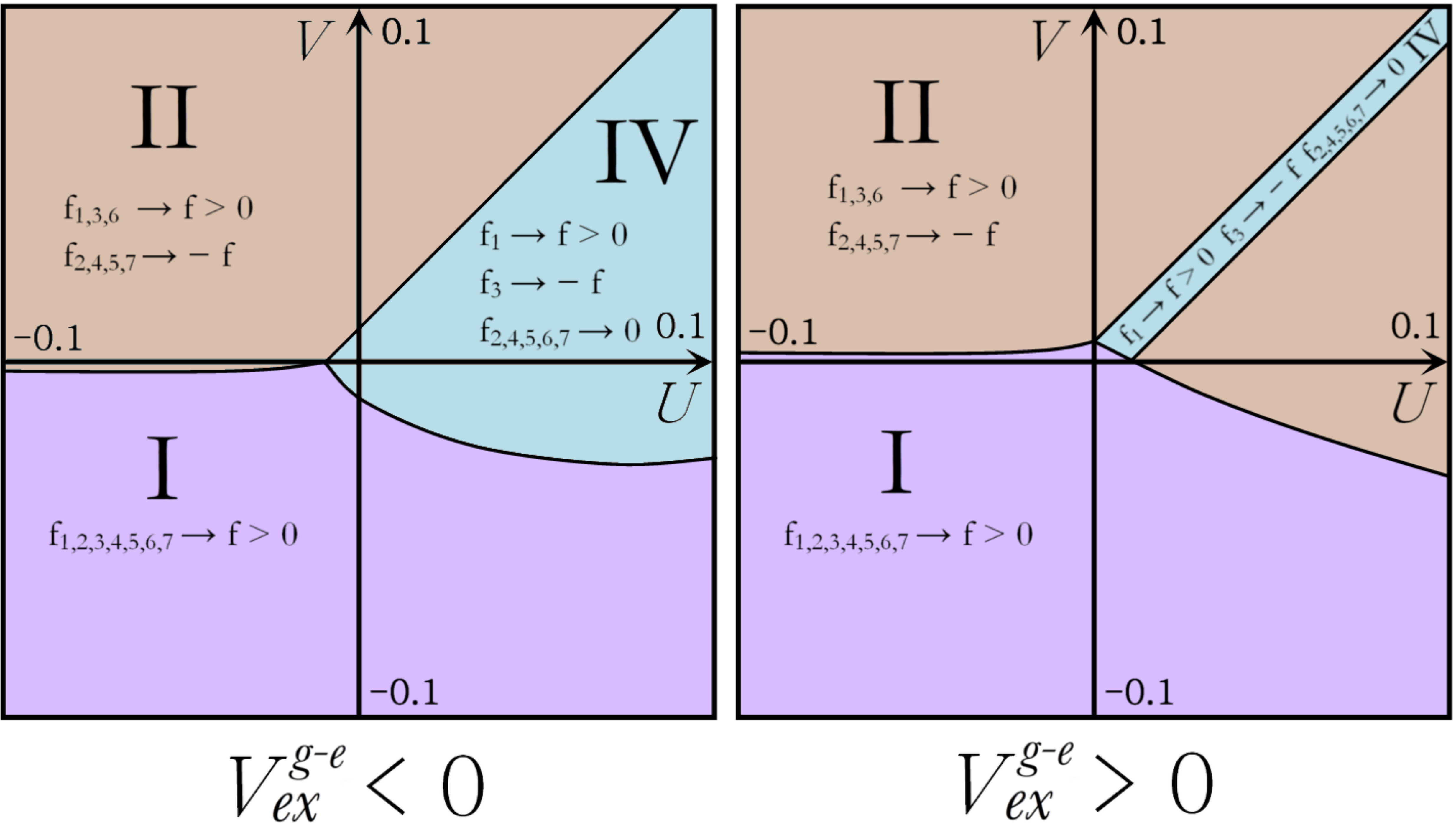}
\caption{Numerical one-loop RG phase diagram for the $g-e$ model (\ref{eqn:Gorshkov-Ham}) 
with $t_g = t_e$ and $U_{gg} = U_{ee} $ ($N>2$).}
\label{fig:PhasediagNRGegsym}
\end{figure}

 We now look at the $g-e$ model. We first consider the isotropic case with $t_g = t_e$ and $U_{gg} = U_{ee} $ 
 so that the one-loop  RG equations are still given by Eqs. (\ref{RGbis}) with the initial conditions (\ref{couplingsgeisotrop}).
The numerical analysis is presented in Fig. \ref{fig:PhasediagNRGegsym} depending on
the sign of $V_{\text{ex}}^{g\text{-}e}$.

On top of the asymptotes (I) and (II) of Eq. (\ref{asymptotRG}) which occur in the $p$-band model, we have
a new one which defines the region IV in Fig. \ref{fig:PhasediagNRGegsym}:
\begin{eqnarray}
&\mbox{IV}& \; :  f_1 = - f_3  = f^{*} > 0, \frac{f_{2,4,5,6,7}}{f_1} \rightarrow 0 .
\label{asymptotRGIV}
\end{eqnarray}

In the orbital anisotropic case $t_g \ne t_e$ and $U_{gg} \ne U_{ee} $, the analysis is much simpler.
Indeed, model (\ref{HamSUN}) describes three commuting marginal SU($N$)$_1$ current-current interactions
with one-loop RG equation of the general form: $\dot{f} = f^2$. 
The fate of these perturbations in the far IR limit depend then only on the sign of the coupling constants
 $g_m$ and $g$.
The different scattering lengths of two atoms collision have been determined recently experimentally
for strontium and ytterbium atoms. \cite{Zhang2014,Scazza2014,Cappellini2014}
The interactions $U_{gg}$ and $ U_{ee}$ are always repulsive so that $\lambda_m <0$ for $m=g,e$ and 
$V_{\text{ex}}^{g\text{-}e} >0$, i.e. $ \lambda < 0$, for ytterbium atoms.  \cite{Scazza2014,Cappellini2014}
Yet in the following, for completeness, we will consider the two cases $V_{\text{ex}}^{g\text{-}e} >0$ and $V_{\text{ex}}^{g\text{-}e} < 0$ as in Fig. \ref{fig:PhasediagNRGegsym}.
When $V_{\text{ex}}^{g\text{-}e} >0$,  the interactions in the effective low-energy Hamiltonian (\ref{HamSUN})  are then marginal irrelevant and a fully-gapless 
$2N$-component Luttinger liquid behavior with central charge $c=2N$ emerges in the IR limit.
In contrast, when $V_{\text{ex}}^{g\text{-}e} < 0$, i.e. $ \lambda > 0$,  the 
last interaction in Eq.  (\ref{HamSUN})  are marginal relevant and opens a 
a spin gap for the nuclear-spin degrees of freedom. The resulting phase is then a gapless $c=2$ phase
which stems from the two-component Luttinger behavior  (\ref{HamU1}) of 
the charge and orbital degrees of freedom.

\section{Phase diagrams}\label{sec:phases}
In this section, we investigate the nature of the dominant electronic instability when $N>2$ in each of the 
corresponding regions of Figs. \ref{fig:PhasediagNRG},\ref{fig:PhasediagNRGegsym}. The zero-temperature phase diagrams 
of the $g-e$ model (\ref{eqn:Gorshkov-Ham}) and the $p$-band model (\ref{pbandmodel}) can then be deduced from
this analysis.

\subsection{Region I}
In region I of Figs. \ref{fig:PhasediagNRG}, \ref{fig:PhasediagNRGegsym}, the one-loop  
RG equations flow along a special ray where $f_i =  f^{*}  > 0$ ($i=1, \ldots, 7$).
The RG Eqs. (\ref{RGbis})  becomes degenerate: $\dot{f} = f^2$ which signals the presence
of an enlarged symmetry. In fact,  in region I of  Fig. \ref{fig:PhasediagNRG}, the Hamiltonian  (\ref{lowenergyham}) enjoys a dynamical symmetry enlargement  \cite{lin,saleur} in the far IR with the emergence of an 
higher SU(2$N$) symmetry which unifies orbital and nuclear-spin degrees of freedom:
\begin{eqnarray}
{\cal H}^{\rm int *}_{\text{so}} =  f^{*}  {\cal J}_L^A {\cal J}_R^A,
   \label{SU2NGN}
\end{eqnarray}
where  ${\cal J}_L^A = (J_\text{L}^a,  j_{\text{L}}^i , J_{\text{L}}^{a,i})$ are left chiral SU(2$N$)$_1$ currents  
with $A = 1, \ldots, 4N^2 -1$.
The IR Hamiltonian takes the form of an SU(2$N$) Gross-Neveu (GN) model \cite{GN} which is an integrable massive
field theory when $ f^{*}  > 0$ \cite{andrei}. The orbital and nuclear spin degrees of freedom are thus fully gapped
and a $c=1$ critical phase is formed which stems from the gapless charge degrees 
of freedom described by the bosonic field $\Phi_c$ in Eq. (\ref{luttbis}). Using the general duality approach to 1D interacting
fermions of Ref. \onlinecite{Boulat-A-L-09}, one expects the emergence of a gapless $2k_F$ charge-density
wave (CDW) phase due to this SU(2$N$)  symmetry enlargement.
Its lattice order parameter is defined through: $n_{2k_F}(j) = \sum_{m,\alpha} e^{i 2 k_F x}  
c^{\dagger}_{m\alpha,j} c_{m\alpha,j}$, with  $x= ja_0$, and $m = g,e$ or $m = p_x, p_y$ respectively
for the $g-e$ and $p$-band models. In the continuum limit (\ref{contlimitdirac}) we get:
\begin{equation}
n_{2k_F} =  R_{m \alpha}^\dagger L_{m\alpha}.
\label{2kFCDW}
\end{equation}
Since this operator is an SU($2N$) singlet, it is worth expressing it
in the non-interacting  U(1)$_c \times$ SU($2N$)$_1$ CFT basis. In this respect,
we use the so-called non-Abelian bosonization
which enables one to find a bosonic description for fermionic bilinears in such a basis \cite{Witten84,Knizhnik-Z-84,affleck86} :
\begin{eqnarray}
L^{\dagger}_{l \alpha} R_{p \beta} \sim \exp \left( i \sqrt{2 \pi /N} \Phi_c \right) g_{p\beta, l \alpha},
\label{nonabelboso}
\end{eqnarray}
where $g$ is the  SU($2N$)$_1$ primary field  with scaling dimension $\frac{2N-1}{2N}$ which 
transforms in the fundamental representation of SU($2N$). \cite{dms}
The IR physics which results from the strong-coupling regime of the SU($2N$) GN model (\ref{SU2NGN})
can then be inferred from a simple semiclassical analysis. In particular,
the interacting part of the SU($2N$) GN model can be expressed in terms of ${\rm Tr} \; g$: 
${\cal H}^{\rm int *}_{\text{so}} \sim  -  f^{*} |{\rm Tr} \; g |^2$, so that $\langle {\rm Tr} \; g \rangle \ne 0$
in the ground state of that phase since $ f^{*} > 0$.
Using the correspondence (\ref{nonabelboso}), we thus obtain in region I
\begin{equation}
n_{2k_F} \sim \exp \left( - i \sqrt{2 \pi K_c /N} \Phi_c \right) ,
\label{2kFCDWRegI}
\end{equation}
where we have rescaled the charge bosonic field by its Luttinger parameter.
The equal-time density-density correlation can then easily be computed as follows:
 \begin{equation}
 \langle n (x)  n (0) \rangle \simeq 
 A  \cos(2 k_F x) x^{-K_c/N}  - \frac{N K_c}{\pi ^2 x^2},
\label{2kfdensitycorrel}
\end{equation}
where $n(x)$ is the continuum description of the lattice density operator $n_i$ and $A$ is 
some non-universal amplitude.

Due to the existence of the SU($N$)$_s$ nuclear-spin symmetry, there is no pairing instability which competes
with the $2k_F$ CDW in region I. Indeed, a general superconducting pairing operator $c^{\dagger}_{m \alpha,i} 
c^{\dagger}_{n \beta,i}$ is not a singlet under the SU($N$)$_s$ symmetry when $N>2$. In the continuum limit,
its nuclear-spin part cannot sustain a non-zero expectation value in the gapful SU(2$N$) invariant model (\ref{SU2NGN}).
All pairing instabilities are thus fluctuating orders with short-ranged correlation functions in stark contrast
to the conclusion of Ref. \onlinecite{Szirmai2013}.
However, we may consider a molecular superfluid (MS) instability made
of $2N$ fermions as in 1D cold fermionic atoms with higher spins \cite{Capponi2008,Roux-C-L-A-09,Lecheminant2005,phle}:
$M^{\dagger}_i = \prod_{\alpha=1}^{N} c^{\dagger}_{g \alpha,i} c^{\dagger}_{e \alpha,i}$ for the $g-e$ model
or $M^{\dagger}_i = \prod_{\alpha=1}^{N} c^{\dagger}_{p_x \alpha,i} c^{\dagger}_{p_y \alpha,i}$  for the $p$-band model. 
In stark contrast to fermionic pairings, $M_i$ is a singlet under  the SU(2$N$) symmetry. The equal-time correlation function of this MS order parameter has been determined in Refs. \onlinecite{Capponi2008,phle} in a phase where the SU($2N$) 
symmetry is restored in the far IR limit as in Eq. (\ref{SU2NGN}):
\begin{equation}
 \langle   M^{\dagger} \left(x\right)  M  \left(0\right) \rangle
 \sim  x^{-N/K_c} .
\label{2Ntets}
\end{equation}
We thus see that $2k_F$ CDW and MS instabilities compete. In particular, 
a dominant MS instability requires $K_c >N$  and thus a fairly large value of $K_c$
which, with only short-range interaction, is not guaranteed. As shown in Refs. \onlinecite{Capponi2008,Roux-C-L-A-09}, 
in the low-density regime and for attractive interactions, 
such value of the Luttinger parameter can be reached for onsite interactions which
signals the emergence of a molecular Luttinger phase. Appart from that case, we expect a dominant
$2k_F$ CDW in region I.

\subsection{Region II}

In region II of Figs. \ref{fig:PhasediagNRG},\ref{fig:PhasediagNRGegsym}, the one-loop  
RG equations flow along the special ray (II) of Eq. (\ref{asymptotRG}).
The resulting interacting IR Hamiltonian takes the form of the SU(2$N$) GN model (\ref{SU2NGN})
when a duality transformation ${\cal D}_{1}$ on the left currents of Eq. (\ref{lowenergyham}) is performed:
\begin{eqnarray}
J_L^{a,(x,y)} &\stackrel{{\cal D}_{1}}{\rightarrow}& - J_L^{a,(x,y)} \nonumber \\
J_L^{a,z} &\stackrel{{\cal D}_{1}}{\rightarrow}& J_L^{a,z}\nonumber  \\
j_L^{(x,y)} &\stackrel{{\cal D}_{1}}{\rightarrow}& - j_L^{(x,y)} \nonumber \\
j_L^{z} &\stackrel{{\cal D}_{1}}{\rightarrow}& j_L^{z} \nonumber \\
J_L^{a} &\stackrel{{\cal D}_{1}}{\rightarrow}& J_L^{a},
\label{dualitytransregionII}
\end{eqnarray}
whereas all right-moving currents remain invariant. This duality symmetry leads to $f_{2,4,5,7} \rightarrow - 
f_{2,4,5,7}$ in Eq. (\ref{lowenergyham}) which is indeed a symmetry of the one-loop RG equations (\ref{RGbis}).
The duality transformation has a simple interpretation on the left-moving Dirac fermions
(the right ones being untouched):
\begin{eqnarray}
L_{1 \alpha} &\stackrel{{\cal D}_{1}}{\rightarrow}& L_{1 \alpha} 
\nonumber \\
L_{2 \alpha} &\stackrel{{\cal D}_{1}}{\rightarrow}&  - L_{2 \alpha},
\label{dualitytransDiracregionII}
\end{eqnarray}
where the orbital index is denoted by $1 \equiv g, p_x$ and $2 \equiv e, p_y$ for
the $g-e$ and $p$-band models  respectively.
Since the SU(2$N$) GN model (\ref{SU2NGN}) is a massive field theory, we deduce that orbital and nuclear-spin
low-lying excitations are fully gapped in region II. Taking account the charge
degrees of freedom (\ref{luttbis}), we conclude that a Luttinger-liquid phase  with central charge
$c=1$ is stabilized. The nature of the dominant electronic instability of that phase
can be deduced from the duality transformation (\ref{dualitytransDiracregionII}) on the $2k_F$-CDW order parameter
(\ref{2kFCDW}), which governs the IR physics in region I:
\begin{eqnarray}
n_{2k_F} =  R_{l\alpha}^\dagger L_{l\alpha} \stackrel{{\cal D}_{1}}{\rightarrow} {\cal O}^{2k_F}_{\rm ODW_z} 
&=& R_{1 \alpha}^\dagger L_{1\alpha} - R_{2 \alpha}^\dagger L_{2\alpha} \nonumber \\ 
&=&  R_{l \alpha}^\dagger 
\sigma^{z}_{lm} L_{m\alpha} .
\label{2kFODW}
\end{eqnarray}
The underlying lattice order parameter is the $2k_F$ component of an orbital density wave (ODW) along the $z$-axis in
the orbital subspace:
\begin{equation}
 {\cal O}_{\rm ODW_z}(i) =  c^{\dagger}_{l \alpha,i} \sigma^{z}_{lm} c_{m \alpha,i} ,
\label{2kFODWlatt}
\end{equation}
with $l,m = g, e$ and $l,m = p_x, p_y$ for respectively the $g-e$ model (\ref{eqn:Gorshkov-Ham})
and the $p$-band model (\ref{pbandmodel}).
Using the result (\ref{2kfdensitycorrel}) in region I, the leading asymptotics of 
the equal-time ODW correlation function is:
\begin{equation}
 \langle {\cal O}_{\rm ODW_z}  \left(x\right)  {\cal O}_{\rm ODW_z}  \left(0\right) \rangle
 \sim  \cos(2 k_F x) x^{-K_c/N} .
\label{2kFODWcorrel}
\end{equation}
Again as in region I, this instability competes with the MS one (\ref{2Ntets}).

\subsection{Region III}

Asymptote (III) of the one-loop RG flow occurs only in the region III of 
the $p$-band model (see Fig. \ref{fig:PhasediagNRG}). Along this special ray,
the interacting IR Hamiltonian is again equivalent to the SU(2$N$) GN model (\ref{SU2NGN})
when a duality transformation ${\cal D}_{2}$ on the left currents of Eq. (\ref{lowenergyham}) is performed:
\begin{eqnarray}
J_L^{a,(z,y)} &\stackrel{{\cal D}_{2}}{\rightarrow}& - J_L^{a,(z,y)} \nonumber \\
J_L^{a,x} &\stackrel{{\cal D}_{2}}{\rightarrow}& J_L^{a,x} \nonumber \\
j_L^{(z,y)} &\stackrel{{\cal D}_{2}}{\rightarrow}& - j_L^{(z,y)} \nonumber \\
j_L^{x} &\stackrel{{\cal D}_{2}}{\rightarrow}& j_L^{x} \nonumber \\
J_L^{a} &\stackrel{{\cal D}_{2}}{\rightarrow}& J_L^{a},
\label{dualitytransregionIII}
\end{eqnarray}
whereas all right-moving currents remain invariant. This duality symmetry leads to $f_{3,4,6,7} \rightarrow - 
f_{3,4,6,7}$ which is indeed a symmetry of the one-loop RG equations (\ref{RGbis}) and 
 expresses as follows in terms of the left-moving Dirac fermions:
\begin{eqnarray}
L_{1 \alpha} &\stackrel{{\cal D}_{2}}{\rightarrow}& L_{2 \alpha} \nonumber \\
L_{2 \alpha} &\stackrel{{\cal D}_{2}}{\rightarrow}&  L_{1 \alpha} ,
\label{dualitytransDiracregionIII}
\end{eqnarray}
the right-moving Dirac fermions being invariant under the transformation.
We have again a gapless $c=1$ phase where orbital and nuclear spin degrees of freedom are fully gapped.
The dominant instability of phase III is obtained from the 
$2k_F$-CDW order parameter (\ref{2kFCDW}) by performing the duality symmetry (\ref{dualitytransDiracregionIII}):
\begin{eqnarray}
n_{2k_F} =  R_{l\alpha}^\dagger L_{l\alpha} \stackrel{{\cal D}_{2}}{\rightarrow}  {\cal O}^{2k_F}_{\rm ODW_x}  &=& 
R_{1 \alpha}^\dagger L_{2 \alpha} + R_{2\alpha}^\dagger L_{1\alpha}  \nonumber \\
&=& 
R_{l \alpha}^\dagger \sigma^x_{l m} L_{m\alpha} .
\label{2kFregionIII}
\end{eqnarray}
The latter being related to the continuum description of the $2k_F$ component of the 
an ODW along the $x$-axis in the orbital subspace:
\begin{equation}
{\cal O}_{\rm ODW_x}(i) =      c^{\dagger}_{l \alpha,i} \sigma^{x}_{lm} c_{m \alpha,i} ,
\label{2kFregionIIIlatt}
\end{equation}
with $l,m = p_x, p_y$.
We thus get a power-law behavior for its equal-time correlation function:
\begin{equation}
 \langle {\cal O}_{\rm ODW_x}  \left(x\right)  {\cal O}_{\rm ODW_x}  \left(0\right) \rangle
 \sim  \cos(2 k_F x) x^{-K_c/N} ,
\label{2kFrungcorrel}
\end{equation}
which competes also with the MS one  (\ref{2Ntets}).

\subsection{Region IV}\label{sec:regionIV}

In the region IV of Fig. \ref{fig:PhasediagNRGegsym} for the $g-e$ model, the one-loop RG flow is attracted
along the special ray (\ref{asymptotRGIV}). Contrary to the previous regions, this asymptote is not
described by an SU($2N$) dynamical symmetry enlargement in the IR limit.
Along the line (\ref{asymptotRGIV}),  the interacting
part of the Hamiltonian density which governs the IR physics of region IV reads as follows:
\begin{eqnarray}
{\cal H}^{\rm int *}_{\text{so}}  &=&
   \frac{\pi  f^{*}}{N} \left( J_L^aJ_R^a  -   2 J_L^{a,z} J_R^{a,z}  \right) \nonumber \\
&=&
 \frac{2 \pi  f^{*}}{N} \left( J_{g L}^a J_{e R}^a  + J_{g R}^a J_{e  L}^a \right),
\label{lowenergyhamregionIV}
\end{eqnarray}
where we have used the SU($N$)$_1$ currents defined in Eq. (\ref{currentaniso}).
Performing a chiral transformation ($\Omega$) 
on the left-moving Dirac fermions $ L_{g, e \alpha} \stackrel{\Omega}{\rightarrow}  L_{e, g \alpha}$,
model (\ref{lowenergyhamregionIV})  takes the form of two independent U($N$) Thirring  (or chiral GN) model, one for
each orbital degrees of freedom:
\begin{eqnarray}
&& {\cal H}^{*}_{\text{so}}  = 
- i v_\text{F} \left(:R_{m \alpha} ^\dag \partial_x R_{m \alpha} ^{\phantom \dag}: - 
  :L_{m \alpha}^\dag \partial_x L_{m \alpha}^{\phantom \dag}: \right)  \nonumber \\
&+& \frac{2 \pi  f^{*}}{N} J_{m L}^a J_{m R}^a = \frac{\pi v_F}{N} \sum_{m=g,e} \left( :J^2_{mL}: + :J^2_{mR}: \right)
 \nonumber \\
&+& \frac{2\pi v_F}{N+1} \left( :J_{m L}^a J_{m L}^a:  + L \rightarrow R \right)
+ \frac{2 \pi  f^{*}}{N} J_{m L}^a J_{m R}^a  ,
\label{Thirringge}
\end{eqnarray}
where $J_{mL,R}$ denotes the U(1) chiral currents for each orbital $m=g,e$ as in 
Eq. (\ref{currentaniso}).
The U($N$) Thirring model is an integrable field theory and for $f^{*}>0$ its non-Abelian
SU($N$) spin sector has a spectral gap. \cite{andrei}
Two gapless U(1) degrees of freedom, one for each orbital, emerges then in the far IR limit, giving rise to 
an extended $c=2$ gapless phase. 

This phase is governed by the competition between two ODW instabilities along the $x$ and $y$ axis:
\begin{eqnarray}
{\cal O}^{2k_F}_{\rm ODW_x}  &=& 
R_{g \alpha}^\dagger L_{e  \alpha} + R_{e  \alpha}^\dagger L_{g \alpha} \nonumber \\ 
& \stackrel{\Omega}{\rightarrow}& R_{m \alpha}^\dagger L_{m  \alpha} 
\nonumber \\
{\cal O}^{2k_F}_{\rm ODW_y}  &=& 
- i R_{g  \alpha}^\dagger L_{e  \alpha} + i R_{e \alpha}^\dagger L_{g \alpha} \nonumber \\ 
& \stackrel{\Omega}{\rightarrow}&
 - i  R_{g  \alpha}^\dagger L_{g  \alpha} + i R_{e \alpha}^\dagger L_{e \alpha} ,
\label{competeODWge}
\end{eqnarray}
after applying the preceeding chiral transformation of the left-moving Dirac fermions. 
We now apply the non-Abelian bosonization approach and use the following identification,
similar to Eq. (\ref{nonabelboso}):
\begin{eqnarray}
L_{m \alpha}^\dagger R_{m \beta} \sim e^{i \sqrt{4 \pi/N} \Phi_{m}}  g_{ \beta \alpha} ,
\label{NonabelbosoSUN}
\end{eqnarray}
where there is no sum over $m$ and $g$ is the SU($N$)$_1$ primary field with scaling dimension $\frac{N-1}{N}$ which acts in the nuclear-spin sector. One finds then the identification:
 \begin{eqnarray}
{\cal O}^{2k_F}_{\rm ODW_x}  &\sim& \left( e^{ - i \sqrt{4 \pi/N} \Phi_{g}} + e^{ - i \sqrt{4 \pi/N} \Phi_{e}}
 \right) {\rm Tr} g^{\dagger}   \nonumber \\
{\cal O}^{2k_F}_{\rm ODW_y}  &\sim& \left( e^{ - i \sqrt{4 \pi/N} \Phi_{g}} - e^{ - i \sqrt{4 \pi/N} \Phi_{e}}
 \right) {\rm Tr} g^{\dagger}  .
\label{competeODWbosege}
\end{eqnarray}
The next step of the approach is to consider symmetric and antisymmetric combination of the bosonic fields:
\begin{equation}
\Phi_{+} = \frac{ \Phi_{g} + \Phi_{e}}{\sqrt 2}, \Phi_{-} = \frac{ \Phi_{g} - \Phi_{e}}{\sqrt 2},
\label{chargeorbitalbosonge}
\end{equation}
and taking into account that $\langle  {\rm Tr} g^{\dagger} \rangle \ne 0$ in the ground state of the Thirring
models (\ref{Thirringge}), we finally get:
\begin{eqnarray}
{\cal O}^{2k_F}_{\rm ODW_x}  &\sim& e^{ - i \sqrt{2 \pi K_+ /N} \Phi_{+}}  \cos \left( \sqrt{2 \pi K_{-}/N} \Phi_{-}\right)
   \nonumber \\
{\cal O}^{2k_F}_{\rm ODW_y}  &\sim& e^{ - i \sqrt{2 \pi K_+/N} \Phi_{+}}  \sin \left( \sqrt{2 \pi K_{-}/N} \Phi_{-}\right),
\label{competeODWfinge}
\end{eqnarray}
where $K_{\pm}$ are the Luttinger parameters for the bosons $\Phi_{\pm}$. 
One can relate these parameters to the charge and orbital ones ($K_{c,o}$) which occur in the 
U(1)$_{g}$ $\times$ U(1)$_{e}$ description of density fluctuations  (\ref{HamU1}) of the anisotropic $g-e$ model. 
The chirality transformation $ L_{g, e \alpha}  \stackrel{\Omega}{\leftrightarrow}  L_{e, g \alpha}$ leads to 
$\Phi_c  \stackrel{\Omega}{\leftrightarrow}  \Phi_+$ and $\Theta_o   \stackrel{\Omega}{\leftrightarrow}  -\Phi_-$, 
so that $ K_+ = K_c$ and 
$K_- = 1/K_o$. The parameter $K_+$ corresponds thus to the charge Luttinger parameter $K_{c}$ which is given by Eq. (\ref{Luttingerparameters}) while $K_- = 1/K_o \simeq 1$ according to our one-loop RG approach in region IV.
Using the definition of the ODW along the $x$-axis (\ref{2kFregionIIIlatt}) and a similar one for the $y$-axis,
the leading asymptotics of the equal-time ODW correlation functions are then given by:
\begin{eqnarray}
\!\!\!\!\!\!\!\!\!\langle {\cal O}_{\rm ODW_x}(x)  {\cal O}_{\rm ODW_x}(0) \rangle
& \sim&  \cos(2 k_F x) x^{-K_c/N - 1/(NK_o)}  \nonumber \\
\!\!\!\!\!\!\!\!\! \langle {\cal O}_{\rm ODW_y}(x)  {\cal O}_{\rm ODW_y}(0) \rangle
& \sim&  \cos(2 k_F x) x^{-K_c/N - 1/(NK_o)} ,
\label{corrODWfinge}
\end{eqnarray}
with  $K_{o} \simeq 1$.
The region IV with extended $c=2$ quantum criticality describes the competition between these two ODW instabilities.

\subsection{Harmonic line of the $p$-band model}

We now consider the harmonic line of the $p$-band model (\ref{pbandmodel}) with $U_1 = 3 U_2 >0$
which plays a special role for repulsive interaction when $N>2$.
Indeed, according to the numerical RG phase diagram, depicted in Fig. \ref{fig:PhasediagNRG}, this harmonic line corresponds to the transition line between phase II and phase III.
It belongs thus to the self-dual manifold of the duality ${\cal D}_{1} {\cal D}_{2}$ symmetry which is defined as:
\begin{eqnarray}
J_L^{a,(x,z)} &\stackrel{{\cal D}_{1} {\cal D}_{2}}{\rightarrow}& - J_L^{a,(x,z)} \nonumber \\
J_L^{a,y} &\stackrel{{\cal D}_{1} {\cal D}_{2}}{\rightarrow}& J_L^{a,y} \nonumber \\
j_L^{(x,z)} &\stackrel{{\cal D}_{1} {\cal D}_{2}}{\rightarrow}& - j_L^{(x,z)} \nonumber \\
j_L^{y} &\stackrel{{\cal D}_{1} {\cal D}_{2}}{\rightarrow}& j_L^{y} \nonumber \\
J_L^{a} &\stackrel{{\cal D}_{1} {\cal D}_{2}}{\rightarrow}& J_L^{a} .
\label{dualitytransharmonicline}
\end{eqnarray}
The self-duality condition gives a strong constraint on the interacting
part of the Hamiltonian density which governs the IR physics
of the harmonic line for repulsive interaction:
\begin{eqnarray}
{\cal H}^{\rm int *}_{\text{so}} =
    g_1 J_L^aJ_R^a   + g_4  J_L^{a,y} J_R^{a,y}  +  g_7  j_L^y j_R^y ,
\label{lowenergyhamharmlinerep}
\end{eqnarray}
where the numerical RG flow gives $f_1 = - f_4 = f^{*} >0$ and $f_7/f_1  \rightarrow 0$ along the 
the harmonic line. 
After performing  a $\pi/2$-rotation along the $x$-axis in the orbital space, the physics along the harmonic line is  
equivalent to that of region IV with central charge $c=2$.
Using the result (\ref{corrODWfinge}), we deduce that the harmonic line, which has an extended U(1)$_o$ symmetry, describes the competition between the ODW orders (\ref{2kFODWcorrel}, \ref{2kFrungcorrel}) along the $z$ and $x$ axis. 
The leading asymptotics of equal-time ODW correlation functions (\ref{2kFODWcorrel}, \ref{2kFrungcorrel}) acquire then
an additional power-law contribution along the $p$-band harmonic line:
\begin{eqnarray}
\!\!\!\!\!\!\!\!\! \langle {\cal O}_{\rm ODW_x}(x) {\cal O}_{\rm ODW_x}  (0) \rangle
& \sim&  \cos(2 k_F x) x^{-K_c/N - 1/(NK_o)}  \nonumber \\
\!\!\!\!\!\!\!\!\! \langle {\cal O}_{\rm ODW_z}(x)  {\cal O}_{\rm ODW_z}(0) \rangle
& \sim&  \cos(2 k_F x) x^{-K_c/N - 1/(NK_o)} ,
\label{corrODWfin}
\end{eqnarray}
with $K_o \simeq 1$. One observes that the two ODW order parameters are unified along the harmonic line. In this respect, the latter plays a similar role as the XY line  in the spin-1/2 antiferromagnetic XYZ Heisenberg chain \cite{giamarchischulz}.

\subsection{Phase diagrams and quantum phase transitions}

With all these results, we can now map out the zero-temperature phase diagram of the $g-e$ model (\ref{eqn:Gorshkov-Ham}) and the $p$-band model (\ref{pbandmodel}) at incommensurate filling.

\subsubsection{$p$-band model}

\begin{figure}[!ht]
\centering
\includegraphics[width=0.5\columnwidth,clip]{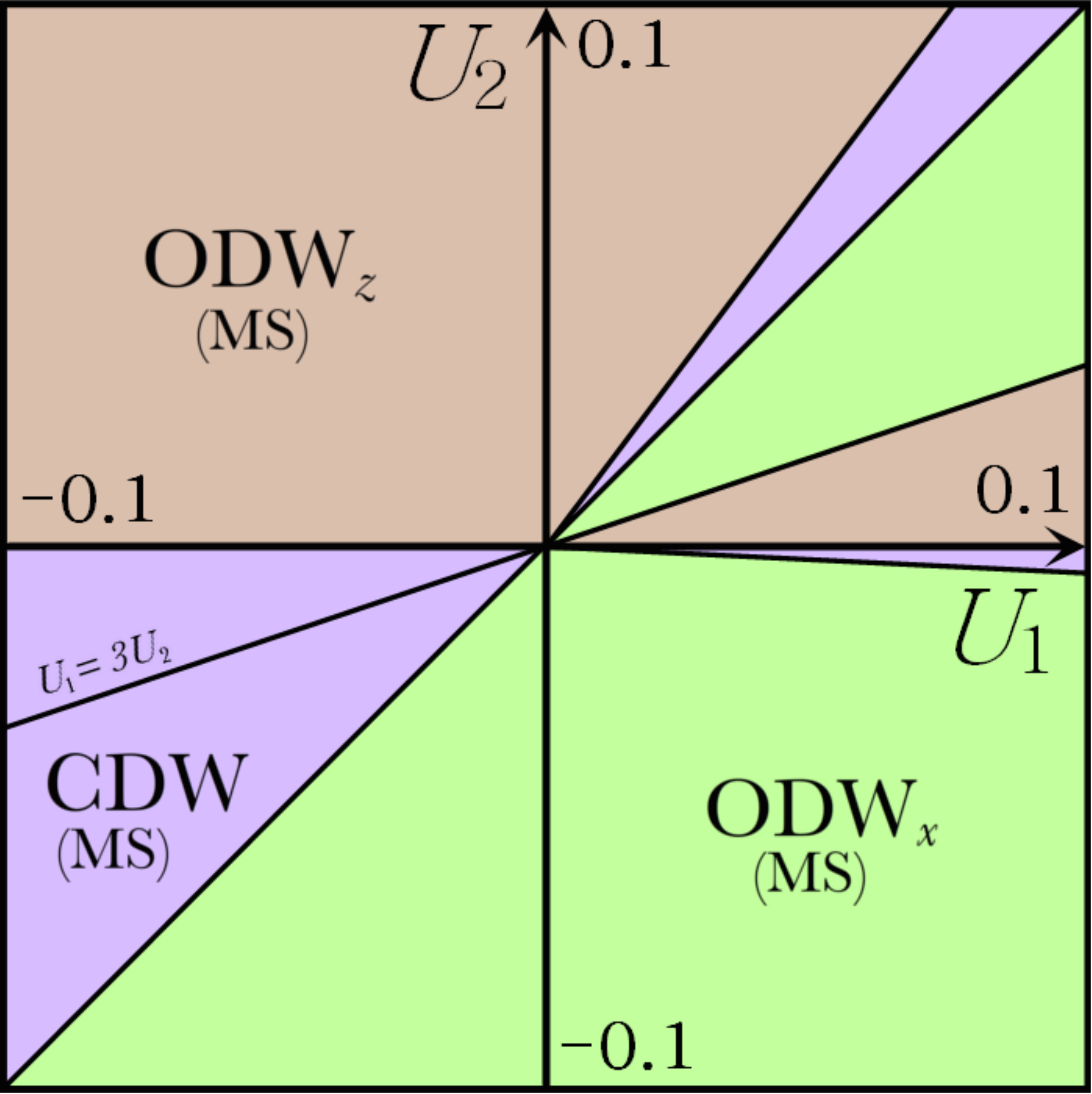}
\caption{Phase diagram  of the $p$-band model (\ref{pbandmodel}) for $N>2$. All phases are gapless with
central charge $c=1$ and the dominant instabilities are indicated together with the competing MS one.}
\label{fig:Phasediagpband}
\end{figure}

Fig. \ref{fig:Phasediagpband} presents the phase diagram of the $p$-band model for $N>2$ deduced from Fig. \ref{fig:PhasediagNRG} and the analysis of dominant instabilities in regions I,II, and III.

The quantum phase transition between the ODW$_x$ and  ODW$_z$ phases belongs to the self-dual manifold of the
${\cal D}_{1} {\cal D}_{2}$ duality symmetry. The repulsive harmonic line $U_1 = 3 U_2 >0$  corresponds to one transition.
The analysis of that line reveals that the transition between  ODW$_x$ and  ODW$_z$ phases is governed by
a $c=2$ CFT with gapless charge and orbital modes.

As seen in Fig. \ref{fig:Phasediagpband}, the transitions between the CDW and ODW$_x$ phases 
occur along the special line $U_1 = U_2$ where the $p$-band model becomes equivalent to the two decoupled single-orbital
SU($N$) Fermi Hubbard model with coupling $U_1$. We then deduce that the $U_1 = U_2 >0$ (respectively 
$U_1 = U_2 < 0$) line has a gapless behavior with central charge $c=2N$ (respectively $c=2$).

The nature of the phase transition  between the CDW and  ODW$_z$ phases can be investigated by 
looking at the self-dual manifold of the duality ${\cal D}_{1}$ symmetry. 
The RG flow is then attracted along the line $ f_1 =  f_3 = f_6 = f^{*} >0$, and $f_2 = f_4 =  f_5 = f_7 = 0$.  The interacting
part of the Hamiltonian density along this self-dual manifold reads as follows:
\begin{eqnarray}
\!\!\! {\cal H}^{\rm int *}_{\text{so}}  \!&=&\!
   \frac{\pi  f^{*}}{N} \left( J_L^aJ_R^a  +  2 J_L^{a,z} J_R^{a,z}  +\frac{2}{N} j_L^zj_R^z \right)  \nonumber \\
\!\!\!\!&=&\!
 \frac{2\pi  f^{*}}{N} \left( J_{p_x L}^a J_{p_x R}^a + J_{p_y L}^a J_{p_y R}^a  +\frac{1}{N} j_L^zj_R^z \right) ,
\label{lowenergyhamregionselfdual}
\end{eqnarray}
where we have used the SU($N$)$_1$ currents defined in Eq. (\ref{currentaniso}) for each orbital $m=p_x,p_y$.
A spin gap is formed for the nuclear-spin degrees of freedom which stems from the marginal relevant SU($N$) current-current
interaction. A gapless $c=2$ behavior then results from the spin-gap opening. This result can
be simply understood from Fig. \ref{fig:Phasediagpband} as we observe  that the $U_2=0$ line with  $U_1<0$ is a transition line between the CDW and  ODW$_z$ phases. The $p$-band model along that line corresponds to 
two decoupled attractive SU($N$) Hubbard models so that a $c=2$ quantum critical behavior is thus expected
in accordance  to the previous self-dual analysis.
 
\subsubsection{Isotropic $g-e$ model}
\begin{figure}[!ht]
\centering
\includegraphics[width=\columnwidth,clip]{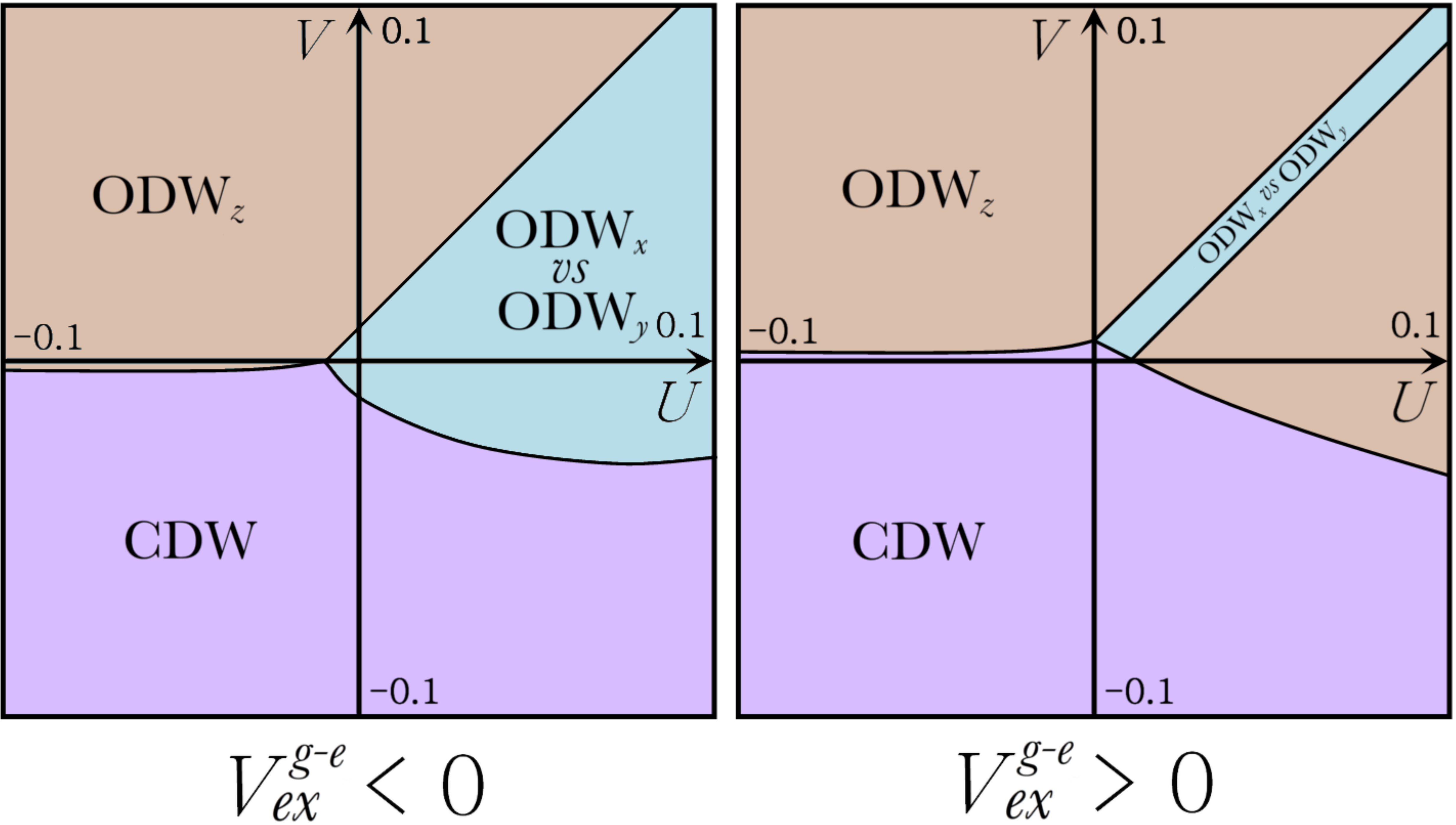}
\caption{Phase diagram for the $g-e$ model (\ref{eqn:Gorshkov-Ham}) 
with $t_g = t_e$ and $U_{gg} = U_{ee} $ ($N<2$).}
\label{fig:Phasediagegsym}
\end{figure}

We now focus to the $g-e$ model (\ref{eqn:Gorshkov-Ham}) 
with $t_g = t_e$ and $U_{gg} = U_{ee} $. The resulting phase diagram is given 
Fig. \ref{fig:Phasediagegsym}.

The nature of the transition between the CDW and  ODW$_z$ phases is identical 
to that of the $p$-band model. We then focus only on the phase transition with $V_{\text{ex}}^{g\text{-}e}>0$ and 
sufficently strong $U>0$, which is relevant to ytterbium atoms.
According to Fig. \ref{fig:Phasediagegsym}, a phase transition between 
ODW$_z$ and the $c=2$ gapless phases might occur. 
The one-loop RG numerical analysis predicts that the transition is governed by 
the manifold with an enlarged SU(2)$_{o}$ orbital rotationnal symmetry with 
low-energy Hamiltonian:
\begin{eqnarray}
{\cal H}^{\rm int *}_{\text{so}}  =
   \frac{\pi  f^{*}}{N} \left( J_L^aJ_R^a  - \alpha J_L^{a,i} J_R^{a,i}  + \beta {\vec j}_L \cdot {\vec j}_R \right) ,
\label{lowenergyhamregionselfdualge}
\end{eqnarray}
where $ f^{*} >0$ and $ 0 < \alpha, \beta <1$ ($ \alpha= \beta= 1/3$  when $N \rightarrow + \infty$). 
The SU($N$)$_2$ sector reaches the strong-coupling behavior before the others
and the SU($N$)$_2$ current-current interaction is a massive integrable model. A mass gap is thus generated
for the nuclear spin degrees of freedom. After integrating out the latter
ones, the analysis of the long-distance physics for the orbital degrees of freedom
is similar to the one in Refs. \onlinecite{Bois-C-L-M-T-15,nonne} and we find the low-energy theory:
\begin{eqnarray}
{\cal H}^{\rm int *}_{\text{o}}  =
 \gamma \; {\rm Tr} {\Phi}^{(1)} +  \frac{\pi  f^{*} \beta}{N}    \; {\vec j}_L \cdot {\vec j}_R ,
 \label{lowenergyhamregionorbitalge}
\end{eqnarray}
where ${\Phi}^{(1)}$ is the spin-1 primary field of the SU(2)$_N$ CFT with scaling
dimension $4/(N+2)$ and $\gamma >0$. The effective Hamiltonian (\ref{lowenergyhamregionorbitalge}) is the low-energy
theory of the spin-$N/2$ SU(2) Heisenberg chain derived by Affleck and Haldane
in Ref. ~\onlinecite{affleckhaldane}.  As shown in Refs. \onlinecite{affleckhaldane,tsvelik}, model (\ref{lowenergyhamregionorbitalge}) has a spectral gap, when $N$ is even,  while it describes a massless flow to the SU(2)$_1$ CFT when $N$ is odd. We thus conclude that the transition is first-order (respectively
critical with $c=1$ behavior) in the $N$ even (respectively odd) case.

\subsubsection{Anisotropic $g-e$ model}

As seen in Sec. \ref{sec:RGge}, the anisotropic $g-e$ model for generic repulsive interactions ($U_{gg} >0, U_{ee}>0,
V_{\text{ex}}^{g\text{-}e}>0$) has critical properties in the $2N$-component Luttinger liquid
universality class. All degrees of freedom are gapless in that phase directly relevant
to ytterbium ultracold atoms since one expects $V_{\text{ex}}^{g\text{-}e}>0$ in that case \cite{Scazza2014,Cappellini2014}.
The leading instability of that phase is an $2k_F$ CDW (\ref{2kfdensitycorrel}) or $2k_F$ SU($N$) spin-density wave (SDW) with continuum order parameter:
\begin{equation}
{\cal S}^A_{2k_F} = R^{\dagger}_{m \alpha} T^{A}_{\alpha \beta} L_{m \beta} ,
\label{SDW}
\end{equation}
with $m=g,e$. 
Using the representation (\ref{NonabelbosoSUN}) and the change of basis (\ref{transU1}), we obtain the leading asymptotics 
of the equal-time density-density correlation function:
\begin{eqnarray}
 \langle n \left(x\right) n \left(0\right) \rangle \sim \sum_{m=g,e} \cos\left(2k_{Fm} x\right)   x^{- \Delta_m}  ,
\label{2kfcdwcorphase}
\end{eqnarray}
with $N \Delta_m = 2 (N - 1) + {\bar v} (K_c + K_o) \frac{K_m}{v_m}$.
We have obtained a similar estimate for the correlations
which involve the $2k_F$ SDW operator.
At this point, these two correlation functions have  the same
power-law decay. The logarithmic corrections will lift this 
degeneracy and we expect that the SDW operator will be the dominant
instability for repulsive interactions as in the SU(2) case \cite{Gogolin-N-T-book,Giamarchi-book}.

The second possible phase for the anisotropic $g-e$ model  is a $c=2$ gapless
phase when $V_{\text{ex}}^{g\text{-}e}<0$. 
As discussed in Sec. \ref{sec:RGge},  there is now a spin-gap for the nuclear-spin degrees of freedom
and the two gapless modes are the bosonic fields $\Phi_c$ and $\Phi_o$ of Eq. (\ref{HamU1}). 
A two-component Luttinger liquid governs then the low-energy properties of this phase.
The physics is very similar to that of Region IV of Sec. \ref{sec:regionIV}  since the formation of the spin gap
in the nuclear-spin sector is described by the same marginal relevant current-current interaction 
of Eq. (\ref{lowenergyhamregionIV}). We deduce that the $c=2$ gapless phase with $V_{\text{ex}}^{g\text{-}e}<0$ corresponds to the competition between the ODWs along the $x$-axis and $y$-axis (\ref{competeODWge}).

\section{Conclusion}

We have presented a detailed field-theory analysis of 1D two-orbital SU($N$) fermionic models 
which governs the low-energy properties of ultracold alkaline-earth like fermionic atoms loading
into a 1D optical lattice. 

Using a formalism which takes explicitly into account the non-Abelian SU($N$)$_s$ symmetry, we have found that the 
U(1)$_c$ charge sector decouples from the orbital-nuclear spin ones at incommensurate filling.
For the $p$-band model, a spectral gap is formed in the latter sector and as a consequence, most of the 
phase diagrams are occupied by Luttinger liquid phases with a single gapless bosonic mode (hence a 
central charge $c=1$). More gapless degrees of freedom can be found along phase transition lines with
U(1)$_{\rm o}$ invariance. In this respect,  the harmonic line $U_1 = 3 U_2 >0$ of the $p$-band model
plays a special role with the existence of two gapless modes.We have shown that this line describes the competition between two different orbital density waves when $N>2$.

The analysis of the $g-e$ SU($N$) model is similar and generically an SU($N$) spin gap for the nuclear-spin
degrees of freedom is formed. However, for repulsive interactions and in the anisotropic case $U_{gg} \ne U_{ee}$,
a situation which is directly relevant to experiments with alkaline-earth like fermionic atoms, 
a fully gapless phase with central charge $c=2N$ occurs with $2N$-component Luttinger liquid physics.

We have clarified the nature of the dominant instability in each phase of 
two-orbital SU($N$) fermionic models. In stark contrast to the $N=2$ case,
no dominant pairing instability are found. The hallmark of these incommensurate two-orbital SU($N$) models
is the competition between various density instabilities when $N>2$. The leading superfluid instability turns out
to be a molecular one found by bound states of $2N$ fermions as in the single-orbital 
attractive1D SU($2N$) fermionic Hubbard model \cite{Capponi2008,Roux-C-L-A-09}.

In the light of the very recent experimental progress with strontium or ytterbium cold fermionic quantum
gases, we hope that it will be possible in the future to unveil part of the rich phase diagram of 
two-orbital SU($N$) models that we describe in this paper.

\section*{Acknowledgements}
The authors are grateful to S. Capponi, M. Moliner, and K. Totsuka for important discussions. We would like to thank CNRS for financial support (PICS grant).




\end{document}